# Undecidability of Adjacent Equality for Insertion, Shuffle, and Crossover Language Operations

Charles E. Hughes
Computer Science Department
University of Central Florida 32751
hughes@ucf.edu

## Abstract

We study a family of language operations based on insertion, shuffle, and crossover and investigate the undecidability of adjacent equality together with finite convergence and associated spectrum questions. Insertion and shuffle operations on formal languages arise in formal language theory, models of concurrency, and biologically inspired computation. This paper studies a different question from the usual closure problem, specifically whether an increasing sequence of languages generated by repeated insertion, or by increasing the permitted degree of bounded shuffle, reaches an instance of adjacent equality after finitely many stages. We show that several such adjacent equality questions are undecidable. In particular, reaching such an adjacent equality event is undecidable for each of the following: iterated insertion of a regular language into a context-free language; bounded shuffle of a regular language with a context-free language as the bound increases; and the corresponding self-insertion and self-bounded-shuffle hierarchies for context-free languages. The new reductions proceed directly from the undecidability of context-free-language universality, using separator-delimited block constructions and, for self-operations, an absorbing regular language of guard violations. Earlier trace-based proofs relied on mortality and uniform halting.

More generally, we investigate finite-stage equality and stabilization (persistent equality) in hierarchies generated by insertion and bounded shuffle. In addition to giving substantially simpler proofs of earlier undecidability results, we obtain general criteria for one-step equality, develop new reductions for self-insertion, and identify several open problems, including structural questions concerning insertion depth and degree whose resolution determines whether adjacent equality necessarily implies permanent stabilization.

## 1. Introduction

For languages A and B over an alphabet Σ, insertion interleaves pieces of one word with pieces of another. Shuffle generalizes insertion by allowing arbitrarily many alternations. The classical literature asks whether language families are closed under these operations and whether equations or containment properties involving them are decidable. Here the central question is instead one of finite adjacent equality, asking the question, if we permit one more insertion stage, or one more insertion cut, does the generated language eventually reach an adjacent stage where no new elements are inserted?

The distinction is important. An operation may be perfectly well defined at every finite stage while no algorithm can determine whether some finite stage introduces any new elements. The results below show this undecidability property exists even when the sets are restricted to context-free languages and interactions between regular and context-free languages. The property of finite stabilization to some fixed set is discussed but, at this point, still left as an open area for further research, although we do identify some promising pathways to address that stronger convergence property. We also introduce the insertion-degree spectrum, prove the **No-Gap Property** for unary languages, and identify a conditional extension from singleton languages to finite languages whose shuffle outputs uniquely determine their source pairs, thereby isolating source-pair ambiguity as a principal obstacle in the general case.

## 2. Definitions and Notations

Let Σ be a finite alphabet and let A, B ⊆ Σ*. For k ≥ 1, define k-insertion by

$A \rhd^{[k]} B = \{ x_1 y_1 x_2 y_2 \ldots x_k y_k x_{(k+1)} \mid y_1 \ldots y_k \in A, x_1 \ldots x_{(k+1)} \in B, \text{and all } x_i, y_j \in \Sigma^* \}$.

When k=1, we write A ⊳ B as simple insertion. If A = B, the operation is called k-self-insertion (or self-insertion for k=1) and $A \rhd^{[k]}$ and A ⊳ are used as a shortened form of $A \rhd^{[k]} A$ and A ⊳ A, respectively. Since any factor $x_i$ or $y_i$ can be the string of length 0 (λ), every concatenation of a word of A with a word of B is obtainable as a k-insertion. Thus, $A \bullet B \subseteq A \rhd^{[k]} B$, and the degree-indexed sequence of sets is non-decreasing under set inclusion, that is, $A \rhd^{[k]} B \subseteq A \rhd^{[k+1]} B$. Insertion closure is defined as $A \rhd^* B = \cup_{k \geq 1} (A \rhd^{[k]} B)$.

The ordinary shuffle A ⧢ B is $A \rhd^* B$, the union over all insertion degrees. The bounded shuffle through degree k is $A ⧢^{[k]} B = \cup_{1 \leq j \leq k} (A \rhd^{[j]} B)$. Since $A \rhd^{[j]} B \subseteq A \rhd^{[j+1]} B$, bounded shuffle reduces to k-insertion. Thus, $A ⧢^{[k]} B = A \rhd^{[k]} B$ and so ⧢ also produces a sequence of non-decreasing sets. Ordinary and bounded self-shuffle use notations $A ⧢^{[k]}$ and A ⧢.

For iterated insertion, write $A^{\rhd[k]}(0)\, B = B$ and $A^{\rhd[k]}(m+1)\, B = A \rhd^{[k]} (A^{\rhd[k]}(m)\, B)$. The analogous notation is used for iterated simple and self-insertion.

Two additional operators of interest are unconstrained crossover, $A \otimes_u B = \{ wz, yx \mid wx \in A \text{ and } yz \in B\}$ and constrained crossover, $A \otimes_c B = \{ wz, yx \mid wx \in A \text{ and } yz \in B, |w| = |y|, |x| = |z| \}$. Self-unconstrained-crossover, self-constrained-crossover, and both iterated crossovers are defined as with other operations. The iterative version of unconstrained crossover produces a non-decreasing series of sets, but the corresponding series produced by iterated constrained crossover can evolve with earlier elements excluded from later sets in the series.

A sequence of insertions, shuffles, or crossovers reach a stage of adjacent equality when two consecutive stages are equal. If the operation has the property that single step equality implies permanent stabilization, then we say the series has reached a fixed point.

## 3. Related Work

The shuffle operation [1] and the more primitive insertion operator [22] have been studied extensively in the computer science literature due to their inherent mathematical interest and their relation to other problems, such as interleaved execution in concurrent systems. More recently, these operations have become of interest in molecular computing, with the proof that contextual insertions and deletions are sufficient to simulate Turing machines, showing the computational completeness of molecular systems based on these two simple operations alone [4][21].

Research addressing issues of closure of classes of languages under the insertion and shuffle operations have a long history. This includes early papers on these operations [4][19][20][21][22][24], with the closure properties of the related deletion operation addressed in [23]. Decidability properties were first considered in [21] for shuffle and in [8] for deletion. A comprehensive early presentation of these topics and a more complete discussion of the notation used here may be found in [15] and [19].

Insertion, deletion, and shuffle operations remain active topics. Recent work by Ibarra and McQuillan [17][18] develops broad classes of insertion operations and proves decidability and undecidability results for language equations and related properties. [18] explicitly covers classical operations ranging from concatenation to shuffle within a generalized insertion framework. [17] studies stable, anti-stable, and error-correctable properties for contextual insertion and deletion.

Shuffle also remains active independently of insertion equations. Barloy, Cadilhac, and Ockerlund [2] study shuffles of context-free languages along regular trajectories and characterize structural conditions under which context-freeness is or is not preserved.

While recent publications are closely related in subject matter, we are unaware of any publications that state the convergence problems proved here: finite convergence of consecutive iterated-insertion stages or consecutive bounded-insertion degrees, either for regular/CFL interaction or for CFL self-interaction.

## 4. Foundational Results

**Single step equality**

In [10], the authors presented a short proof that one cannot decide, for an arbitrary context free language L, whether $L \bullet L = L$. Here, we present a generalization of this result that was first shown in [14] and extended in [15]. This generalization provides a proof of the undecidability of single step equality of any class of languages where membership is decidable, universality is undecidable, and the operator subsumes concatenation.

**Theorem 1.** Let $\mathbb{C}$ be a class of languages where, for any L in $\mathbb{C}$, where L is over the alphabet $\Sigma$, the membership question, is $\Sigma \cup \{\lambda\} \subseteq L$, is decidable, but the universality problem, is $L = \Sigma^*$, is undecidable. Moreover, let $\odot$ be a binary operator over sets in $\Sigma^*$, such that for $A, B \subseteq \Sigma^*$, $A \odot B \subseteq \Sigma^*$ and suppose $L \bullet L \subseteq L \odot L$, then the problem of deciding whether $L \odot L = L$ is at least as hard as universality

*Proof*. Let L be an arbitrary language in $\mathbb{C}$. We claim that $L = \Sigma^*$ iff $L \odot L = L$.
Clearly, if $L = \Sigma^*$ then $\Sigma^* = L \bullet L \subseteq L \odot L \subseteq \Sigma^*$ and so $L \odot L = \Sigma^*$.
Since $\Sigma \cup \{\lambda\} \subseteq L$, $L^* = \Sigma^*$. If $L \odot L = L$ and $L \bullet L \subseteq L \odot L$ then $L = L^* = \Sigma^*$.
The consequence is $L = \Sigma^*$ iff $L \odot L = L$. QED

**Corollary.** The operators denoted as simple insertion ($\rhd$), ordinary shuffle (ш), and unconstrained crossover ($\otimes$u) satisfy all the criteria of Theorem 1 and, hence, the problems of deciding whether $L \rhd L = L$, L ш L = L, and $L \otimes u\ L = L$ are each undecidable.

Note that Constrained crossover ($\otimes$c) does not meet the subsumption criterion. As it is not necessarily the case that $L \bullet L \subseteq L \otimes c\ L$ and therefore is not addressed by this theorem. Thus, the complexity of the question, “does $L \otimes c\ L = L$?”, is open.

Other interesting properties of these operators are, for CFLs, A and B,

- $A \bullet B$ is a CFL
- $A \rhd B$ is a CFL
- $A \rhd_{[k]} B$ is not necessarily a CFL, for k>1
- A ш B is not necessarily a CFL [2].

Additionally, there are open complexity questions to be addressed, such as:

- For CFL, L, what is the computational complexity of “does $L \otimes c\ L = L$”?

**Mortality and uniform halting time**

A Turing machine is mortal if it halts from every starting configuration, not merely from a conventional start state on a finitely marked input. A machine has uniform (constant) running time if there exists an integer s such that every finitely described starting configuration halts within s steps. Theorems 2 and 3 below were both first shown in [13] based on the 1966 mortality results of Hooper [9].

**Theorem 2.** For the Turing-machine model that allows infinite starting configurations proposed by Hooper [9], mortality is equivalent to the existence of a uniform finite bound on running time. The resulting ConstantRunningTime set of Turing-machines is recursively enumerable but not recursive.

**Proof sketch.** If a uniform bound exists, mortality is a direct consequence. Conversely, suppose every finite configuration halts but no uniform bound exists. Organize configurations that must be fully scanned before the Turing machine ends up in a finitely branching tree. Unbounded finite behavior yields an infinite branch by König's Infinity Lemma, producing an infinite configuration on which the machine runs forever. Hence mortality implies a uniform bound. Recursive enumerability follows by testing all configurations of bounded size for candidate bounds; undecidability follows from mortality.

**Adjacent equality**

Let # be an operator on formal languages that, when iterated, produces a monotonically non-decreasing series. Consistent with iterated insertion**,** $L^{\#}(0) = L$ and $L^{\#}(k+1) = L^{\#}(L^{\#}(k))$. We say that #, when applied to L, reaches adjacent equality iff
$\exists n\ [\ L^{\#}(n+1) = L^{\#}(n)\ ]$. If $\exists n\ \forall k \geq n\ [\ L^{\#}(k+1) = L^{\#}(k)\ ]$, we say # on L has attained permanent stabilization

**Theorem 3.** For an arbitrary context-free language L, the problem of deciding whether there exists a finite n>0 such that $L^n = L^{(n+1)}$, equivalently reaches adjacent equality, is undecidable. Here $L^k$ is shorthand for $L^{\bullet}(k)$, where $\bullet$ is the concatenation operator.

**Proof sketch.** The reduction encodes computation traces so that longer powers correspond to longer valid traces. A uniform bound on computation length is shown equivalent to finite convergence of the power sequence. Details are presented in [13]. Note that, for concatenation, adjacent equality implies permanent stabilization but that is not necessarily true for other operations we address in this paper, so the somewhat weaker result stated here is appropriate.

Given the generalization we saw in Theorem 1, one is tempted to feel this carries over in Theorem 3 to all operators that subsume concatenation, but it is easy to show operators such as $L^{\#}(k) = \Sigma^*$, unusual as this may be, for which convergence to a fixed point occurs after one stage. Fortunately, iterated insertion and iterated shuffle are not such edge cases as we will see later, but that begs the question as to what property might replace subsumption of concatenation, $L \bullet L \subseteq L \otimes L$, in order for us to have a general theorem rather than specific constructions for each individual operator for which undecidability holds.

Theorem 3 is the conceptual ancestor of the proofs about to be presented. Concatenation, used there, is replaced by insertion and shuffle, and a second hierarchy is obtained by increasing the insertion degree rather than the number of iterations.

## 5. Iterated Insertion: Regular into Context-Free Languages

This result was previously shown in [14] by encoding Turing-machine traces. Later variants contain much shorter reductions directly from universality (equality to $\Sigma^*$) of context-free languages [15]. The shorter argument is given here.

**Theorem 4.** It is undecidable, for a regular language R and a context-free language L, whether there exists a $k \geq 0$ such that $R^{\triangleright}(k)\, L = R^{\triangleright}(k+1)\, L$.

**Proof.** Let $L' \subseteq \Sigma^*$ be an arbitrary context-free language. We reduce the undecidable question, is $L' = \Sigma^*$, to the theorem's adjacent equality question? Membership of the string of length zero ($\lambda$) in a CFL is decidable; if $\lambda$ is not in L', then L' is already known not to be $\Sigma^*$. Hence consider the remaining instances with $\lambda$ in L'. Introduce a separator # not in $\Sigma$ and define $L = (L'\#)^*L'$ and $R = \Sigma^*$. Then L is context free and R is regular.

Every word of L is a sequence $w_1\#w_2\#...\#w_n$ in which every block $w_i$ belongs to L'. Because words of R contain no #, a single insertion cannot create, remove, or cross a separator; it can alter only one block. Starting from a word of L, after k ordinary insertion stages, at most k blocks can fail to belong to L'. Conversely, since $\lambda$ belongs to L', any chosen collection of at most k blocks not in L' can be realized by starting with empty blocks and inserting the desired elements of $\Sigma^*$, one per insertion stage. Thus, $R^{\triangleright}(k)\, L$ is exactly the set of separator-delimited strings having at most k blocks outside L'.

If $L' = \Sigma^*$, there are no #-separated blocks outside L' and insertion adds nothing, so convergence is immediate. If L' is not $\Sigma^*$, there exists some non-empty string w in $\Sigma^* - L'$. For every k >= 0, the word consisting of k+1 copies of w separated by # has k+1 non-L' blocks. This set of blocks belongs to
$R^{\triangleright}(k+1)\, L$, by starting with k+1 empty blocks ($\#^k$), but it cannot belong to $R^{\triangleright}(k)\, L$. Hence every stage produces a set strictly larger than its predecessor. Therefore $L' = \Sigma^*$ iff there exists $k \geq 0$ such that
$R^{\triangleright}(k)\, L = R^{\triangleright}(k+1)\, L$. Since universality of context-free languages is undecidable, so is finite adjacent equality. QED

Note that in our construction, adjacent equality occurs when $R^{\triangleright}(1)\, L = L$, but the proof provides a more general result, showing undecidability for the existence of any finite k>0.

## 6. Bounded Shuffle of Regular with Context-Free Languages

The same separator invariant gives an almost identical proof when the parameter is insertion degree rather than iteration depth.

**Theorem 5.** It is undecidable, for a regular language R and a context-free language L, whether there exists a degree n such that $R \sqcup\!\sqcup^{[n]} L = R \sqcup\!\sqcup^{[n+1]} L$; equivalently, whether bounded insertion reaches adjacent equality at a finite degree ($R \triangleright^{[n]} L = R \triangleright^{[n+1]} L$).

**Proof.** As before, choose an arbitrary context free language $L' \subseteq \Sigma^*$, a separator $\# \notin \Sigma$, $L = (L'\#)^*L'$, and $R = \Sigma^*$. A k-shuffle splits one selected word of R into k pieces and inserts those pieces at k positions in a word of L. Since no word from R contains #, k-insertion can affect at most k separator-delimited blocks. Conversely, any prescribed set of at most k bad blocks can be produced by beginning with empty L'-blocks, concatenating the desired replacement words from one word of R, and splitting it at the corresponding boundaries before insertion. Consequently $R \triangleright^{[k]} L$ is exactly the set of separator-delimited strings with at most k blocks outside L'.

If L' = Σ*, the bounded insertion hierarchy stabilizes immediately. If L' is not Σ*, there exists some word w outside L'. The word w#w#...#w with k+1 copies belongs to $R \rhd^{[k+1]} L$ but not to $R \rhd^{[k]} L$. Thus, no finite degree stabilizes. Hence L' = Σ* iff there exists k >= 0 such that $R \rhd^{[k]} L = R \rhd^{[k+1]} L$. Since bounded shuffle through degree k equals k-insertion in this setting, the same argument proves undecidability of adjacent equality for bounded shuffle. QED

## 7. Self-Insertion of a Context-Free Language

The separator proof in Theorem 4 can be adapted to self-insertion, but one extra device is needed. In the regular/context-free construction, the inserted language R = Σ* contains no separator, so a single insertion can affect only one block. Under self-insertion, an inserted word may itself contain separators. We prevent those unintended interactions from contributing new valid words by using a regular insertion-ideal of guard violations, that is

Let L' ⊆ Σ* be an arbitrary context-free language, and, as in Theorem 4, restrict attention to instances with λ ∈ L'. Introduce three symbols @, &, and # not in Σ, and let Γ = Σ ∪ {@, &, #}. Define

T = @ (L'#)* L' &

and let G be the regular language consisting of words having at least two @ symbols, at least two & symbols, some symbol before an @, or some symbol after an &. Equivalently,

$G = \Gamma^*@\Gamma^*@\Gamma^* \cup \Gamma^*\&\Gamma^*\&\Gamma^* \cup \Gamma^+@\Gamma^* \cup \Gamma^*\&\Gamma^+$.

Finally define L = Σ* ∪ T ∪ G. Since Σ* and G are regular and T is context free, L is context free. The important property of G is that it is an insertion ideal: once one of its four guard violations occurs, inserting any additional word cannot remove it. Hence inserting a word of G into any word, or inserting any word into a word of G, always yields another word of G.

For m ≥ 0 let $T_m$ be the set of words $@w_1\#w_2\#...\#w_r\&$ ($r \geq 1$, $w_i \in \Sigma^*$) having at most m blocks $w_i$ outside L'. Thus $T_0$ = T.

**Theorem 6.** It is undecidable, for a context-free language L, whether there exists m such that $L^{\rhd}(m) = L^{\rhd}(m+1)$.

**Proof.** Let L, L', T, and G be defined as above. We claim by induction on m that $L^{\rhd}(m) = \Sigma^* \cup G \cup T_m$, where $L^{\rhd}(0) = L$ and $L^{\rhd}(m+1) = L \rhd (L^{\rhd}(m))$. The claim is immediate for m = 0.

Assume $L^{\rhd}(m) = \Sigma^* \cup G \cup T_m$. If a word of Σ* is inserted into a word of Σ*, the result remains in Σ*. If a word of Σ* is inserted into a word of $T_m$, no @, &, or # is introduced by the inserted word, so at most one separator-delimited block can be changed; the result is in $T_{(m+1)}$. Conversely, any word of $T_{(m+1)}$ that is not already in $T_m$ can be obtained from a word of $T_m$ by replacing one bad block by λ and then inserting that block, as a word of Σ*, at the corresponding position.

If a word of T is inserted into a word of $T_m$, the result contains two @ symbols and two & symbols and is therefore in G. If a word of T is inserted into a word of Σ*, then either the target word is λ, in which case the result is the original word of T, or some target symbol occurs before @ or after &, placing the result in G. All interactions involving G remain in G by the insertion-ideal property. These cases prove the induction.

If L' = Σ*, then every block is good, so $T_m = T_0$ for every m and the self-insertion sequence stabilizes immediately. If L' ≠ Σ*, choose a nonempty w ∈ Σ* − L'. For every m ≥ 0, the word @w#w#...#w& having m+1 copies of w belongs to $T_{(m+1)}$ but not to $T_m$. Hence every stage is strictly larger than its predecessor. Therefore L' = Σ* iff there exists m such that $L^{\rhd}(m) = L^{\rhd}(m+1)$. Since universality of context-free languages is undecidable, finite adjacent equality of self-insertion is undecidable. QED

## 8. Self-Bounded-Shuffle of a Context-Free Language

The same guarded construction also yields a short proof for increasing the degree of self-insertion. The only additional interaction to check is the case in which a structured word of T is split into k pieces and inserted into a word of $\Sigma^*$.

**Theorem 7.** It is undecidable, for a context-free language L, whether there exists n such that $L ⧢^{[n]} = L ⧢^{[n+1]}$; equivalently, whether the self-bounded-shuffle hierarchy achieves adjacent equality.

**Proof.** Use the same languages L', $\Sigma^*$, T, G, L and Tm as in Theorem 6. Since bounded shuffle through degree k equals k-insertion, it is enough to show that $L \rhd^{[k]} = \Sigma^* \cup G \cup Tk$.

If the inserted word is from $\Sigma^*$ and the target is from T, its k pieces contain no guard or separator symbols, so they can affect at most k blocks. Conversely, any prescribed collection of at most k bad blocks can be created from a word of T by beginning with λ in those blocks, concatenating the desired bad block-words into one word of $\Sigma^*$, and splitting that word into the corresponding pieces. Thus, this interaction produces exactly Tk.

Insertion of $\Sigma^*$ into $\Sigma^*$ remains in $\Sigma^*$. Any interaction involving G remains in G. Inserting a word of T into a word of T produces at least two @ and two & symbols and therefore lies in G.

It remains to consider a k-insertion of a word of T into a target word of $\Sigma^*$. If the result is not in G, the unique @ from the T word must be the first symbol of the result and its unique & must be the last. Consequently, all target pieces lying outside the interval delimited by @ and & must be empty. Only target pieces inserted between two consecutive pieces of the split T word can modify T-blocks, and there are at most k−1 such internal boundaries. Hence every non-garbage result of this interaction lies in $T_{(k-1)}$, which is already contained in $T_k$.

Therefore $L \rhd^{[k]} = \Sigma^* \cup G \cup T_k$. If L' = $\Sigma^*$, then $T_k = T_0$ for every k, so the hierarchy is constant. If L' ≠ $\Sigma^*$, choose $w \in \Sigma^* - L'$. The word @w#w#...#w& with k+1 copies of w belongs to $T_{(k+1)}$ but not $T_k$. Thus $L \rhd^{[k]} \neq L \rhd^{[k+1]}$ for any k. Again, since universality of context-free languages is undecidable, adjacent equality problem for self-bounded shuffle is also undecidable. QED

## 9. Finite Permanent-Stabilization

**Corollary.** The finite permanent-stabilization problem is undecidable for each of the hierarchies in Theorems 4–7.

**Proof.** In each reduction, the hierarchy is constant when L' = $\Sigma^*$ and strictly increasing at every stage when L' ≠ $\Sigma^*$. Thus, the constructed hierarchy reaches adjacent equality iff it permanently stabilizes. Since universality of context-free languages is undecidable, this shows finite permanent-stabilization is also undecidable in these four cases.

## 10. Iteration-Depth Spectrum Theorem

Define $i_{A,B}(w) = \min \{m : \exists k \geq 1 \ \& \ w \in A^{\rhd[k]}(m) B\}$ and $\mathcal{I}(A, B) = \cup_{k \geq 1, m \geq 0} \{i_{A,B}(w) : w \in A^{\rhd[k]}(m) B\}$.

**Conjecture (Iteration-Depth No-Gap Conjecture).**
For arbitrary sets of strings A, B, $r \in \mathcal{I}(A, B) \Rightarrow \{0,\ldots,r\} \subseteq \mathcal{I}(A, B)$. Equivalently, $\mathcal{I}(A, B)$ is an initial interval of the non-negative integers.

**Theorem 8.** For arbitrary languages, A, B, each of finite cardinality, $\mathcal{I}(A, B)$ has no gaps. Moreover, if A and B are both non-empty, $\mathcal{I}(A, B) = \mathbb{N}_0$.

**Proof.** The edge cases are: if $A = \varnothing$ or $A = \{\lambda\}$ and $B \neq \varnothing$ , then $\mathcal{I}(A, B) = \{0\}$; and if $B = \varnothing$ then $\mathcal{I}(A, B) = \varnothing$. Both cases satisfy the Insertion-Depth No-Gap Conjecture.

For arbitrary non-empty language S, define $M(S) = \{\max|w| : w \in S\}$.
Let $a = M(A) = \max\{|w| : w \in A\}$ and $b = M(b) = \max\{|w| : w \in B\}$. Since insertion preserves length additively, inserting u into v, regardless of insertion degree (k), produces a word of length $|u|+|v|$. of Choose a word $w_m$ of maximum length produced in stage m. its length, $M(A^{\triangleright[k]}(m)\ B)$, is $b + m*a$, for all $k \geq 1$. Every earlier stage has strings of maximum length $b + j*a$, where $a,b > 0$ and $j < m$, and so $b + j*a < b + m*a$, meaning the m-th stage is the earliest one containing $w_m$, and so $i_{A,B}(w_m) = m$. Therefore, $\mathcal{I}(A, B) = \mathbb{N}_0$.

Hence the Iteration-Depth No-Gap Conjecture holds for all finite language pairs.

## 11. Insertion Degree Spectra and the No-Gap Conjecture

For languages $A, B \subseteq \Sigma^*$, and $w \in A ⧢ B$, define the insertion degree of w relative to A, B by

$d_{A,B}(w) = \min \{ k \geq 1 : w \in A \triangleright^{[k]} B \}$

Define the insertion-degree spectrum $\mathcal{D}(A, B) = \{ d_{A,B}(w) : w \in A ⧢ B \}$.

Since $A \triangleright^{[k]} B \subseteq A \triangleright^{[k+1]} B$, an apparently natural question is whether the attainable minimum degrees can contain gaps.

**Conjecture 1 (No-Gap Conjecture).**

For arbitrary languages A, B, $r \in \mathcal{D}(A, B) \Rightarrow \{1,\ldots,r\} \subseteq \mathcal{D}(A, B)$. Equivalently, $\mathcal{D}(A, B)$ is an initial interval of the positive integers.

$A \triangleright^{[k]} B = A \triangleright^{[k+1]} B$ means that degree k+1 is absent from the spectrum. If the No-Gap Conjecture holds, no larger degree can occur either. Therefore $A \triangleright^{[k]} B = A ⧢ B$, where $A ⧢ B = \cup_{j\geq 1} (A \triangleright^{[j]} B)$.

**Proposition 1. (Conditional on No-Gap Conjecture).**

If $D(A, B)$ is an initial interval of the positive integers and, for some $k \geq 1$, $A \triangleright^{[k]} B = A \triangleright^{[k+1]} B$ then $A \triangleright^{[k]} B = A ⧢ B$.

**Proof.** If two consecutive stages are equal, no word has minimum insertion degree k+1. By the initial-interval property, no word can have minimum insertion degree greater than k. Hence every word of $A ⧢ B$ already belongs to $A \triangleright^{[k]} B$. The reverse containment follows from the definition of shuffle. Therefore $A \triangleright^{[k]} B = A ⧢ B$.

Thus, No-Gap would establish, for arbitrary bounded-insertion hierarchies,

*Adjacent Equality ⟹ Permanent Stabilization*

However, even the singleton case appears nontrivial. For fixed words, $u, v \in \Sigma^*$, define $\mathcal{D}(u, v) = \mathcal{D}(\{u\}, \{v\})$. A shuffle of u and v may admit multiple assignments of its positions in the two source words. Consequently, the minimum insertion degree of a resulting word need not equal the number of insertion runs in any given representation. This representation ambiguity prevents the immediate argument of successively merging insertion runs.

**Theorem 9.** (Unary-Source Singleton Spectrum). Let $u=a^m$ with $m \geq 0$ and let $v \in \Sigma^*$. Let s be the number of symbols of v that are not a. Then $\mathcal{D}(u,v) = \{1\}$ if m=0, and $\mathcal{D}(u,v) = \{1,\ldots,\min(m,s+1)\}$ if m>0. Consequently, the Singleton No-Gap Conjecture holds whenever the inserted word is over a unary alphabet.

**Proof.** Decompose v into maximal a-blocks separated by non-a symbols. If there are s non-a symbols, there are s+1 a-blocks. Every shuffle distributes the m inserted a's among these blocks. Within any block, all inserted a's may be regarded as one insertion run, so the minimum insertion degree is exactly the number of occupied blocks. Any r with $1 \leq r \leq \min(m,s+1)$ is achieved by occupying exactly r blocks. Hence $\mathcal{D}(u,v)=\{1,\ldots,\min(m,s+1)\}$ for m>0. If m=0 then $u=\lambda$ and the only attainable insertion degree is 1. When s=0 this immediately yields $\mathcal{D}(a^m,v)=\{1\}$, recovering the unary singleton case. QED

**Conjecture 2 (Singleton No-Gap Conjecture).**

For $u, v \in \Sigma^*$, $\mathcal{D}(u,v)$ is an initial interval of the positive integers.

**Source-pair ambiguity.** The principal difficulty in extending the singleton conjecture to finite languages is that a word in $A ⧢ B$ may have shuffle representations arising from more than one pair of source words.

Recall that $d_{A,B}(w) = \min \{ k \geq 1 : w \in A \rhd^{[k]} B \}$ and $\mathcal{D}(A, B) = \{ d_{A,B}(w) : w \in A ⧢ B \}$.

Consequently, even if every singleton spectrum $\mathcal{D}(u, v)$ is an initial interval, it does not follow immediately that $\mathcal{D}(A, B)$ is the union of those singleton spectra as, for a given word, $w \in A ⧢ B$, more than one source pair, u, v, may give different values, the smallest of which is the minimum insertion degree.

**Definition.** A pair of languages (A, B) is **source-pair-separable** if distinct source pairs have disjoint shuffle sets, that is, whenever $(u, v) \neq (u', v')$ are in $A \times B$, $(u ⧢ v) \cap (u' ⧢ v') = \emptyset$.

**Proposition 2.** If the Singleton No-Gap Conjecture holds, then the No-Gap Conjecture holds for every source-pair-separable pair of finite languages A and B.

**Proof.** The result is immediate if either language is empty. Otherwise, source-pair separability implies that every $w \in A ⧢ B$ has a unique source pair $(u, v) \in A \times B$. Hence $\mathcal{D}(A, B) = \cup \{\mathcal{D}(u, v) \mid (u, v) \in A \times B\}$. Assuming Singleton No-Gap, each $\mathcal{D}(u, v)$ is an initial interval of the positive integers. Since $A \times B$ is finite, their union is also an initial interval. Therefore $\mathcal{D}(A, B)$ has no gaps. QED

**Finite representation erasure.** Proposition 2 isolates source-pair ambiguity as the obstruction to extending Singleton No-Gap directly to arbitrary finite languages. Even if every $\mathcal{D}(u, v)$ is an initial interval, competing source pairs can lower the minimum insertion degree assigned to particular shuffle words. The remaining question is whether such representation erasure can remove every witness of an intermediate degree while leaving a witness of some larger degree.

**Problem 1** (Finite Representation-Erasure Problem). Suppose A and B are finite languages and every singleton spectrum $\mathcal{D}(u, v)$, for $(u, v) \in A \times B$, is an initial interval. Can competing source-pair representations produce a gap in $\mathcal{D}(A, B)$? A negative answer would show that Singleton No-Gap implies No-Gap for all finite languages.

**Finite-to-arbitrary passage.** There is also a natural compactness question concerning the passage from finite to arbitrary languages. For any fixed shuffle output word w, all source pairs (u,v) satisfying $w \in u ⧢ v$ obey $|u| + |v| = |w|$. Thus $d_{A,B}(w)$ is determined by a finite length-bounded portion of $A \times B$, even when A and B are infinite. This does not immediately reduce the general No-Gap Conjecture to the finite case. However, the occurrence of a degree is witnessed by a single word, whereas the absence of a degree asserts that no such witness occurs in $A ⧢ B$.

**Problem 2** (Finite-to-Arbitrary No-Gap). If the No-Gap Conjecture holds for all finite languages, does it follow that it holds for arbitrary languages?

**Research hierarchy.** The results and questions above suggest the progression
unary languages (proved) → singleton languages → source-pair-separable finite languages → arbitrary finite languages → arbitrary languages.
The first class satisfies No-Gap by Theorem 9. Conditional on Singleton No-Gap, Proposition 2 establishes the property for source-pair-separable finite languages. Whether the remaining implications hold is open. These questions separate two sources of difficulty -- ambiguity among different interleavings of a fixed pair of words, already present in the singleton problem, and ambiguity among different source pairs, which first appears at the language level.

Note that the iteration-depth spectrum discussed in Section 10 has a global monotone invariant (maximum word length), whereas the insertion-degree spectrum has no comparable invariant because changing insertion degree does **not** change total length. That contrast leads us to believe that the insertion-degree conjecture for finite languages is hard for a structural reason, not just because no one has found the right proof.

## 12. Factor Replacement Systems with Residue as a Trace Model

Factor Replacement Systems might be a cleaner starting point, but their non-deterministic application creates a challenge. However, the use of a deterministic trace model based on Factor Replacement Systems with Residues (FRS-R) overcomes this challenge, offering what could be a new, more compact set of proofs.

FRS-R were introduced in class notes by the author in [15] and used extensively in [16]. These systems represent state by a natural number and use unordered rules of the form $ax+b \rightarrow cx+d$. By including residues, these systems can test non-divisibility, thereby overcoming the ordering needed when ordinary FRS systems simulate computational models such as Register Machines. Encoding of INC and DEC register-machine instructions and maintaining state are straightforward. Such systems can be designed to converge to 0, whenever the register machines halts. Building on this register machine encoding, the notes [16] demonstrate that FRS-R computations can simplify the proofs appearing in the quotient-of-CFL convergence construction presented in [12].

The insight from the above is that FRS-R may provide a cleaner numerical source model for trace arguments related to CFLs. However, the analogue of the mortality/constant-time theorem has not been shown for FRS-R models of computation. This suggests an interesting question, does FRS-R have an undecidable uniform-halting property? If so, it could yield a substantially shorter trace-based proof of the concatenation result [13], analogous to the simplification using FRS-R systems and guarded constructions demonstrated for CFL quotient in [16].

## 13. Finite Convergence as a Dynamical Property

Noting that the insertion and a related deletion operation are used in biomolecular computing and dynamical systems; shuffle is used in analyzing concurrency as the arbitrary interleaving of parallel events; and crossover is used in genetic algorithms, leads to some possibly interesting alignments with research outside computability theory.

Specifically, the results above can be viewed as statements about discrete dynamical processes on languages. Starting from a set X0, repeated local interaction may generate a non-descending sequence $X_0 \subseteq X_1 \subseteq X_2 \subseteq \ldots$ . The permanent-stabilization question asks whether $\exists k\ [\ X_m = X_k\ \forall m \geq k\ ]$. For an iterative sequence based on a many-one mapping, $X_{m+1} = F(X_m)$, the equality $X_k = X_{k+1}$ already implies permanent stabilization. For degree-indexed hierarchies such as bounded insertion, whether adjacent equality has this consequence in general is precisely the issue raised by the No-Gap Conjecture.

In the insertion construction of Theorems 4-7, a particularly simple progress measure is available. Words are decomposed into separator-delimited blocks, and one may count the blocks that fail to belong to the source context-free language. A single ordinary insertion can affect at most one block, while a k-insertion can affect at most k blocks. When the source language is universal there are no bad blocks; when it is not universal, arbitrarily many bad blocks can be created and each finite stage can be separated from the next. Thus undecidability arises not from difficulty in computing an individual finite stage, but from the impossibility of deciding whether the entire monotone evolution eventually reaches a fixed point.

This perspective suggests possible connections beyond formal-language theory. Mortality and uniform-halting arguments have long served as reduction mechanisms in discrete, piecewise-affine, and hybrid dynamical systems [5][6]. The present results do not establish undecidability for any additional dynamical-system model; rather, they provide a compact finite-stabilization pattern that may be applied to systems generated by repeated local interaction, rewriting, population interaction, or constrained interleaving.

A useful general question is therefore, for which monotone operators F, is it decidable whether the orbit $X_0$, $F(X_0)$, $F^2(X_0)$, ... reaches a fixed point after finitely many steps? The insertion and bounded-shuffle results show that strong syntactic restrictions on the finite stages need not make this eventual convergence question decidable.

## 14. Open Problems and a Call for Collaboration

The purpose of revisiting these results is more than archival. The author welcomes alternative proofs, stronger results, counterexamples to proposed extensions, historical references that may have been missed, and applications of the finite-convergence construction to other computational or dynamical models. Collaboration is particularly welcome on the questions below.

**General Theory**

- Identify a general bounded-propagation or rank condition that subsumes the separator proofs of Theorems 4 to 7 and yields undecidability of finite convergence for a broad family of operators.
- Determine the exact recursion-theoretic or arithmetical-hierarchy complexity of the finite-convergence problems, rather than undecidability alone.
- Identify restricted language families or restricted insertion/shuffle mechanisms for which finite convergence becomes decidable.
- Study quantitative variants in which convergence is guaranteed and determine whether the least stabilization index can be computed or effectively bounded.

**No-Gap Program**

- Determine whether Insertion-Degree No-Gap for all finite languages implies Insertion-Degree No-Gap for arbitrary languages or identify a counterexample showing that the finite-to-arbitrary implication fails.
- Resolve the Singleton Insertion-Degree No-Gap Conjecture and determine whether Singleton Insertion-Degree No-Gap implies Insertion-Degree No-Gap for all finite languages by ruling out finite representation erasure.
- Determine whether Insertion-Depth No-Gap for all finite languages implies Insertion-Depth No-Gap for arbitrary languages or identify a counterexample showing that the finite-to-arbitrary implication fails.

**Applications**

- Investigate finite stabilization for discrete, piecewise-affine, hybrid, rewriting, population, and evolutionary systems whose dynamics are generated by repeated local interactions.
- Recast the reductions using Factor Replacement Systems with Residue [10],[11], and identify the weakest mortality or uniform-halting property required of the source model.
- Study constrained crossover as it creates interesting theoretical challenges not addressed here.

The open problems above motivate interesting paths forward, such as generalizing the existing proofs, extending the theorems to new formal models, and discovering applications outside formal-language theory. Researchers who recognize an equivalent theorem under different terminology, or who see a natural model to which the reduction can be transferred, are encouraged to pursue the connection.

## Acknowledgements and Historical Comments

The original work [14] on this set of problems was motivated by Masami Ito's November 2004 UCF lecture [19][20] and by earlier work on the finite power property of context-free languages [13]. Ferucio Tiplea, University of Iași, Romania, listened to early versions of the arguments, encouraged the work, and suggested significant simplifications to several proofs and presentations.

The underlying finite-power line of work began with all proofs in place by 1977 except for the connection needed to complete the uniform-halting argument. The author sought suggestions from several colleagues, while insisting that the missing steps be no longer than one page. Stanley Selkow later recognized the key idea after attending a conference in Montreal, where Jaroslav Opatrny of Concordia suggested the connection. That observation and the subsequent construction completed the proof in [13] that was critical to the current work.

## Declaration of Generative AI and AI-assisted Technologies in the Manuscript Preparation Process

During the preparation of this work, the author used OpenAI's ChatGPT for critical discussion of proofs, exploration of alternative proof strategies and possible generalizations, literature-search assistance, identification of potential gaps and counterexamples, and suggestions concerning organization and presentation. AI-assisted discussions cleaned up and simplified many of the proofs appearing in the paper, contributed to the development and examination of the simplified separator and guarded constructions, and initiated the formulation and investigation of the insertion-degree-spectrum and No-Gap questions. The author provided the foundational research including the relationship to universality, and the bases for guarded constructions. The author also posed the iteration-depth No-Gap conjecture, proving it for finite languages, and showed the iteration -degree No-Gap property is true for singleton alphabets. The author also proposed many of the open problems and the potential connections to other models and disciplines. The author independently reviewed, verified, and revised all AI-assisted material and takes full responsibility for all mathematical statements, proofs, citations, and conclusions in this article.